\documentclass[
 superscriptaddress,
 reprint,
 nofootinbib,
 amsmath,
 amssymb,
 aps, 
 prl,
 showkeys
]{revtex4-2}

\usepackage{graphicx}
\usepackage{dcolumn}
\usepackage{bm}
\usepackage{hyperref}

\usepackage{bm}
\usepackage[T1]{fontenc}
\usepackage{inputenc}
\usepackage{upgreek,babel,calc,mathrsfs,textgreek,bbm}
\usepackage{mathtools,graphicx,xcolor}
\usepackage{cleveref}

\crefname{equation}{Eq.}{Eqs.}
\Crefname{equation}{Equation}{Equations}
\crefname{figure}{Fig.}{Figs.}
\Crefname{figure}{Figure}{Figures}

\begin{document}
    \title{
        \textbf{Sparse Interactions Reshape Stability in Random Lotka-Volterra Dynamics} 
    }

    \author{Mattia Tarabolo}
        \email{mattia.tarabolo@polito.it}
        \affiliation{Institute of Condensed Matter Physics and Complex Systems, Department of Applied Science and Technology, Politecnico di Torino, C.so Duca degli Abruzzi 24, I-10129 Torino, Italy}
    
    \author{Luca Dall'Asta}
        \affiliation{Institute of Condensed Matter Physics and Complex Systems, Department of Applied Science and Technology, Politecnico di Torino, C.so Duca degli Abruzzi 24, I-10129 Torino, Italy}
        \affiliation{Italian Institute for Genomic Medicine (IIGM) and Candiolo Cancer Institute IRCCS, str. prov. 142, km 3.95, I-10060 Candiolo (TO), Italy}
        \affiliation{INFN, Turin Via Pietro Giuria 1, I-10125 Turin, Italy}
    
    \author{Roberto Mulet}
        \affiliation{Group of Complex Systems and Statistical Physics, Department of Theoretical Physics, Faculty of Physics. University of Havana, CP 10400, La Habana, Cuba}
    
    \date{\today}
    
    \begin{abstract}
        Classical approaches to ecological stability rely on fully connected interaction models, yet real ecosystems are sparse and structured--a feature that qualitatively reshapes their collective dynamics. Here, we establish a thermodynamically exact stability phase diagram for generalized Lotka-Volterra dynamics on sparse random graphs, resolving how finite connectivity and interaction heterogeneity jointly govern ecosystem resilience. Using a small-coupling expansion of the dynamic cavity method, we derive an effective single-site stochastic process that is solvable via population dynamics. Our approach uncovers a topological phase transition--driven purely by the finite connectivity structure of the network--that leads to multi-stability. This instability is fundamentally distinct from the disorder-driven transitions induced by quenched randomness of the couplings. Our framework overcomes the considerable computational cost of direct simulations, offering a scalable and versatile analysis of stability, biodiversity, and alternative stable states in realistic, large-scale ecological ecosystems.
    \end{abstract}
    
    \maketitle
    The relationship between complexity and stability in large ecological communities has been a central question in theoretical ecology since May's seminal work \cite{mayWillLargeComplex1972, mayStabilityComplexityModel2019, mccannDiversityStabilityDebate2000}.  A widely adopted framework to address this question is the generalized Lotka-Volterra (gLV) model, which describes the coupled population dynamics of interacting species\cite{mayTheoreticalEcologyPrinciples2007} using interaction coefficients that are often treated as quenched random variables \cite{buninEcologicalCommunitiesLotkaVolterra2017,biroliMarginallyStableEquilibria2018,altieriPropertiesEquilibriaGlassy2021,altieriEffectsIntraspecificCooperative2022}.

    Most analytical progress has been achieved in the \emph{fully connected} limit. Here, the large number of interactions per species justifies central-limit arguments, leading to effective Gaussian noise and mean-field behavior \cite{buninEcologicalCommunitiesLotkaVolterra2017,gallaDynamicallyEvolvedCommunity2018}. In this regime, Dynamical Mean-Field Theory (DMFT) and path-integral approaches map the many-body problem onto a single effective degree of freedom, providing a detailed description of fixed points, extinction patterns, and stability boundaries \cite{riegerSolvableModelComplex1989,gallaDynamicallyEvolvedCommunity2018,biroliMarginallyStableEquilibria2018}. However, real ecological networks--from food webs to mutualistic communities--are typically \emph{sparse}, characterized by finite connectivity that defies standard mean-field approximations \cite{dunneNetworkStructureBiodiversity2002,dunneFoodwebStructureNetwork2002,jordanoInvariantPropertiesCoevolutionary2003,thebaultStabilityEcologicalCommunities2010,pichonTellingMutualisticAntagonistic2024}.

    Despite their empirical relevance, a general theoretical framework for the \textit{non-equilibrium} dynamics of gLV models on sparse graphs remains elusive. Recent large-connectivity expansions \cite{aguirre-lopezHeterogeneousMeanfieldAnalysis2024,parkIncorporatingHeterogeneousInteractions2024,poleyInteractionNetworksPersistent2025} provide valuable insights but lose accuracy when the connectivity is finite, as they fail to capture strong local fluctuations. Exact methods for sparse graphs have been proposed, but with significant restrictions. For instance, Ref.~\cite{tonoloGeneralizedLotkaVolterraModel2025} employs a static cavity method that accounts for finite connectivity but is strictly limited to equilibrium states. Conversely, Ref.~\cite{metzDynamicalMeanFieldTheory2025} successfully applies population dynamics to the full non-equilibrium process, but is restricted to \textit{directed} networks where response functions vanish. A very recent study \cite{machadoLocalEquationsGeneralized2025} addresses the undirected case with demographic noise using a similar cavity expansion on single realizations. It relies on a first-order approximation that tracks only the mean abundance, a simplification that proves effective for uncorrelated couplings, but neglects the response functions and self-consistent correlations that may be crucial for a general description of sparse undirected systems. Finally, direct numerical simulations of the dynamics become computationally very demanding as the system size increases, making it difficult to disentangle finite-size effects from genuine phase transitions in the thermodynamic limit. Consequently, the general case of undirected sparse networks--where non-vanishing response functions introduce complex memory effects--remains to be explored.

    To overcome these limitations, we adapt the \emph{Gaussian Expansion Cavity Method} (GECaM) \cite{taraboloGaussianApproximationDynamic2025} to the specific constraints of ecological dynamics. This framework generalizes the extended Plefka expansion \cite{braviExtendedPlefkaExpansion2016} to sparse topologies by performing a small-coupling expansion of the dynamic cavity equations. This procedure maps the interacting $N$-body problem onto a set of decoupled single-species stochastic processes driven by colored noise and self-consistent fields. These fields, which encode the dynamical moments of neighboring species, are determined through a message-passing scheme on the interaction graph, thereby retaining the full heterogeneity of the network structure. By solving these effective equations with a population dynamics algorithm, we averaged both topology and disorder, obtaining the first stability phase diagram for gLV systems with finite connectivity in the thermodynamic limit.
    
    We consider a community of $N$ species whose abundances $x_i(t)$ evolve on a sparse undirected random graph with adjacency matrix $A=\{a_{ij}\}$. The dynamics follows the generalized Lotka-Volterra equations
    \begin{equation}
        \frac{d {x}_i(t)}{dt} = x_i(t) \left[1 - x_i(t) + \sum_{j \in \partial i} J_{ij} x_j(t) \right],
        \label{eq:Lotka-Volterra_eq}
    \end{equation}
    where we have set the intrinsic growth rates and carrying capacities to unity. The sum runs over the set of neighbors $\partial i$ defined by the graph structure. The latter is a quenched realization, drawn from an uncorrelated random graph ensemble with a given degree distribution $p_{\rm deg}(k)$.
    
    The interaction coefficients $J_{ij}$ quantify the effect of species $j$ on $i$. For each interacting pair $(i, j)$, the couplings are drawn from a bivariate Gaussian distribution with mean $\overline{J_{ij}} = m/K$, variance $\mathrm{Var}(J_{ij}) = \sigma^2/K$, and correlation $\mathrm{Corr}(J_{ij}, J_{ji}) = \gamma$. These parameters control the macro-ecological state: $m$ sets the average cooperation or competition, $\sigma$ determines the heterogeneity strength, and $\gamma \in [-1, 1]$ tunes the reciprocity, interpolating between predator-prey dominated ($\gamma \to -1$) and symmetric competitive/mutualistic regimes ($\gamma \to 1$).

    \begin{figure*}
        \centering
        \includegraphics[scale=1]{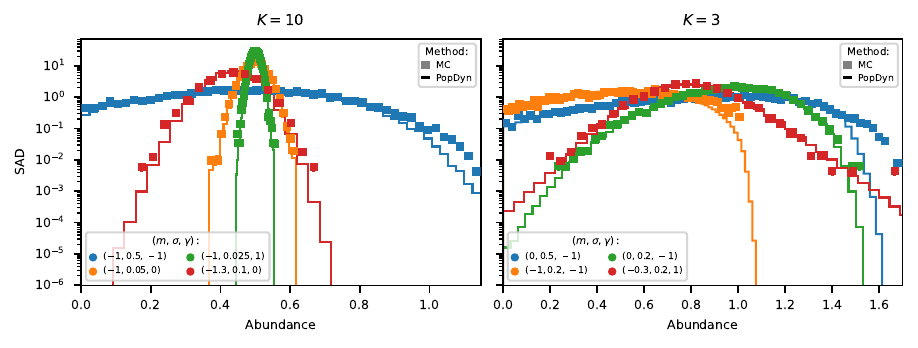}
        \caption{\textbf{Species abundance distributions on sparse random graphs.} Validation of the effective theory against microscopic simulations. The plots compare stationary abundances from direct Monte Carlo integration (MC, markers) with the fixed-point population-dynamics solution (PopDyn, lines). \textbf{Left:} Random Regular Graphs (RRG) with connectivity $K=10$, exhibiting near-Gaussian statistics. \textbf{Right:} Low connectivity $K=3$, showing strong skewness. Parameter sets $(m,\sigma,\gamma)$ are listed in the legends. MC data aggregate 20 graph realizations (20 initial conditions each); error bars denote standard errors. PopDyn population size is $P\in[10^5,10^7]$.}
        \label{fig:SAD_comparison}
    \end{figure*}

    The effective dynamical process is derived using the Gaussian Expansion Cavity Method (GECaM)~\cite{taraboloGaussianApproximationDynamic2025}. Starting from the Martin-Siggia-Rose-Janssen-De Dominicis (MSRJD) path-integral formalism~\cite{martinStatisticalDynamicsClassical1973,janssenLagrangeanClassicalField1976,dedominicisDynamicsSubstituteReplicas1978}, we perform a small-coupling expansion of the resulting dynamic cavity equations. This procedure decouples the interacting system into a set of effective single-site Langevin equations. For a given edge $(i,j)$, the effective abundance $x_{i \setminus j}(t)$ of species $i$ in the absence of species $j$ evolves as
    \begin{widetext}
        \begin{equation}
            \frac{d {x}_{i\setminus j}(t)}{dt} = x_{i\setminus j}(t)\Biggl[1-x_{i\setminus j}(t)+\sum_{k\in\partial i\setminus j}J_{ik}\mu_{k\setminus i}(t) + \sum_{k\in\partial i\setminus j}J_{ik}J_{ki}\int_{0}^{t}dt'R_{k\setminus i}(t,t')x_{i\setminus j}(t') + h_{i\setminus j}(t)+\xi_{i\setminus j}(t)\Biggr],\label{eq:eff_proc_cav}
        \end{equation}
    \end{widetext}
    where $h_{i\setminus j}(t)$ is a probing field and $\xi_{i\setminus j}(t)$ is a Gaussian colored noise with zero mean and correlation $\langle \xi_{i\setminus j}(t)\xi_{i\setminus j}(t')\rangle = \sum_{k\in\partial i\setminus j}J_{ik}^{2}C_{k\setminus i}(t,t')$.
    The dynamics is driven by three self-consistent cavity fields: the mean abundance $\mu_{k\setminus i}(t) = \langle x_{k\setminus i}(t)\rangle$, the response function $R_{k\setminus i}(t,t') = \delta \langle x_{k\setminus i}(t)\rangle / \delta h_{k\setminus i}(t')$, and the connected correlation $C_{k\setminus i}(t,t')$. Here, $\langle \cdot \rangle_{k\setminus i}$ denotes the average over the realization of the stochastic process \cref{eq:eff_proc_cav} for the edge $(k,i)$. These fields encapsulate the memory and feedback effects characteristic of sparse undirected topologies. A detailed derivation of the effective cavity equations is provided in the Supplemental Material~\cite{supp}.
    
    Solving these non-Markovian equations in the time domain is computationally demanding~\cite{royNumericalImplementationDynamical2019}. To proceed, we assume the system relaxes to a time-translation invariant stationary state independent of initial conditions.
    In this limit, the cavity fields become static: $\mu_{i\setminus j}(t) \to \mu_{i\setminus j}^{\star}$, $C_{i\setminus j}(t, t') \to q_{i\setminus j}^{\star}$, and the integrated response $\int dt' R_{i\setminus j}(t,t') \to \chi_{i\setminus j}^\star$. The fluctuating noise $\xi_{i\setminus j}$ freezes into a static random variable $\xi_{i\setminus j}^\star$ with variance $\Sigma_{i\setminus j}^2 = \sum_{k\in\partial i\setminus j} J_{ik}^2 q_{k\setminus i}^\star$.
    Setting $\dot{x}=0$ in \cref{eq:eff_proc_cav}, it gives  the fixed-point solution
    \begin{align}
        x_{i\setminus j}^{\star}(s)&=\frac{\Sigma_{i\setminus j}(\Delta_{i\setminus j}-s)+h_{i\setminus j}}{\Gamma_{i\setminus j}} \nonumber \\
        &\qquad \times \Theta\left(\frac{\Sigma_{i\setminus j}(\Delta_{i\setminus j}-s)+h_{i\setminus j}}{\Gamma_{i\setminus j}}\right),\label{eq:FP_sol_cavity}
    \end{align}
    where the noise is parametrized by $\xi^\star = -\Sigma_{i\setminus j} s$ with $s \sim \mathcal{N}(0,1)$ and we used the short-hand notation for $\Delta_{i\setminus j}=(1+\sum_{k}J_{ik}\mu_{k \setminus i}^{\star})/\Sigma_{i\setminus j}$ and $\Gamma_{i\setminus j}=1-\sum_{k}J_{ik}J_{ki}\chi_{k \setminus i}^{\star}$.
    Taking moments over the Gaussian variable $s$ closes the self-consistency loop, yielding a set of message-passing equations for the static fields $(\mu_{i\setminus j}^\star, q_{i\setminus j}^\star, \chi_{i\setminus j}^\star)$ defined on a single realization of the disorder. While these equations can be solved for any fixed graph instance, the typical behavior of the system in the thermodynamic limit can be extracted employing a population dynamics algorithm (PopDyn), which iterates a large ensemble of cavity fields to sample the stationary distribution over the graph and coupling ensembles (see SM for the details of the message-passing update rules and the algorithmic implementation~\cite{supp}). The code and data are available at~\cite{repo_rGLV}.
    
    \begin{figure}[ht]
        \centering
        \includegraphics[scale=1]{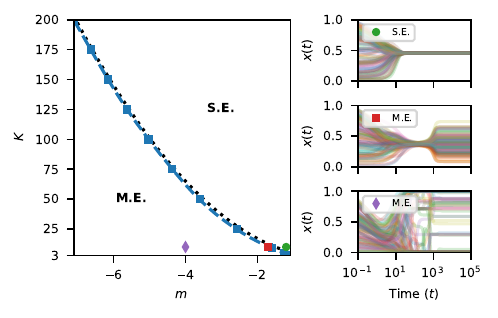}
        \caption{\textbf{Topological transition at $\sigma=0$ on Random Regular Graphs.} Phase boundary in the $(m,K)$ plane for homogeneous couplings ($J_{ij}=m/K, \sigma=0$). \emph{Markers} indicate the critical point $m_c(K)$ where the Population Dynamics (PopDyn) develops a non-vanishing variance $q$, signaling the transition from a globally stable Single Equilibrium (SE) to a fragmented Multiple Equilibria (ME) phase. The solid line has been obtained by interpolation. The \emph{dashed line} plots the analytical stability boundary derived from the Kesten-McKay spectral edge, $m_c(K) = -K/(2\sqrt{K-1})$, marking the loss of linear stability for the fully feasible equilibrium. \textbf{Right panels:} Representative microscopic trajectories of species abundances $x_i(t)$ illustrating the dynamical regimes. Top ($m=-1.2$, SE): All species converge to a unique, homogeneous abundance value. Middle ($m=-1.7$, ME): Just below the transition, species settle into distinct fixed points depending on initial conditions, breaking the homogeneity. Bottom ($m=-4$, deep ME): Deep in the fragmented phase, the dependence on initial conditions persists and the fraction of extinct species ($x_i=0$) increases significantly.}
        \label{fig:sigma0_phase_diagram}
    \end{figure}

    \begin{figure*}[ht]
        \centering
        \includegraphics[scale=1]{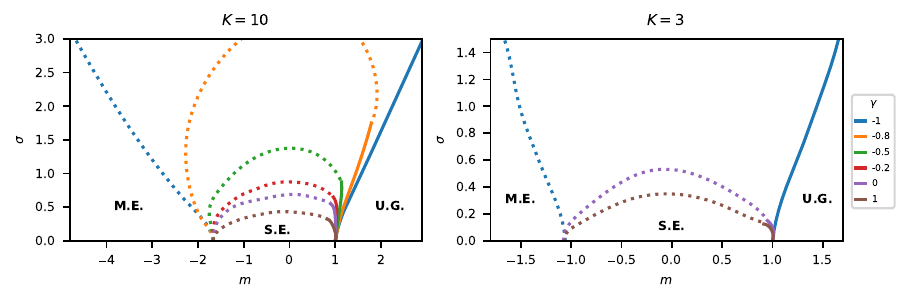}
        \caption{\textbf{Stability phase diagrams in the $(m, \sigma)$ plane.} 
        Results for Random Regular Graphs with $K=10$ (\textbf{left}) and $K=3$ (\textbf{right}) obtained via Population Dynamics. 
        \emph{Dashed lines} mark the transition from the Single Equilibrium (SE) to the Multiple Equilibria (ME) phase; \emph{solid lines} mark the transition to the Unbounded Growth (UG) phase. 
        Different colors correspond to interaction symmetry values $\gamma \in [-1, 1]$ (see legend). 
        Note the shrinking of the stable SE region as $\gamma \to 1$ and the severe compression of stability at low connectivity ($K=3$). 
        Dynamical regimes are identified by the PopDyn behavior: SE (convergent to $q=0$), ME (non-convergent/oscillatory, $q>0$), UG (divergent $\mu$).}
        \label{fig:phasediagram}
    \end{figure*}

    The effective theoretical framework is validated against direct Monte Carlo (MC) simulations of the microscopic dynamics. \Cref{fig:SAD_comparison} displays the stationary Species Abundance Distributions (SADs) for Random Regular Graphs (RRGs) across different model parameters. For relatively high connectivity $K=10$, the SADs appear nearly Gaussian and the population‑dynamics (PopDyn) results show excellent agreement with MC data--including subtle asymmetries. As connectivity is reduced to $K=3$ strong deviations from Gaussianity emerge due to enhanced local fluctuations: the distributions broaden significantly and develop pronounced skewness. In particular, even in this highly heterogeneous regime, where conventional mean‑field approximations typically break down, the PopDyn solution remains in quantitative agreement with microscopic simulations across the entire parameter range tested.

    The structural effects of sparseness can be isolated by setting the strength of the disorder $\sigma=0$, so that the system is governed solely by the mean interaction $m$ and connectivity $K$. A distinctive feature of sparse ecosystems is the emergence of a \emph{topological phase transition} from a single globally stable equilibrium (SE) to a Multiple Equilibria (ME) phase. This transition, absent in fully connected models and first observed numerically in Ref.~\cite{marcusLocalCollectiveTransitions2022}, is identified within our cavity framework as the point where the distribution of the variance $q$ is no longer concentrated at zero. Our derivation explicitly assumes that the system relaxes to a unique fixed point independent of initial conditions; this implies that the stationary state of the effective process is deterministic, and thus its variance must vanish ($q=0$). For $m$ above a critical threshold $m_c(K)$,the population dynamics confirms this hypothesis, relaxing to a unique solution where the entire population satisfies $q=0$. As $m$ drops below $m_c(K)$, the assumption of a unique attractor breaks down. In this fragmented regime, the effective process develops a non-trivial variance ($q>0$), reflecting the dependence on initial conditions characteristic of a landscape with multiple locally stable attractors. Consequently, the population dynamics iterations typically fail to converge to a stationary distribution, exhibiting persistent oscillations or divergence. Although the fixed-point theory cannot be used to quantitatively describe the interior of this phase, the breakdown of the solver accurately marks the stability boundary. 
    \Cref{fig:sigma0_phase_diagram} confirms this picture: the transition coincides perfectly with the theoretical stability bound $m_c(K) = -K/(2\sqrt{K-1})$~\cite{kestenSymmetricRandomWalks1959,mckayExpectedEigenvalueDistribution1981}, which marks the point where the homogeneous, fully feasible equilibrium loses linear stability due to the spectral edge of the adjacency matrix crossing unity. This result establishes that the instability is purely topological, driven by combinatorial exclusion constraints rather than random heterogeneity: when effective competition exceeds a threshold, loops in the sparse graph frustrate global coexistence, forcing the community to fragment.
    
    The study of the full phase diagram in the $(m, \sigma)$ plane reveals the interplay between this topological mechanism and disorder-driven instability. Three distinct phases can be identified, depending on the convergence properties of the population dynamics. In the \emph{Single Equilibrium} (SE) phase, PopDyn converges to a unique fixed point where the distribution of the cavity variance $q$ is concentrated at zero, corresponding to a stable deterministic attractor. In the \emph{Multiple Equilibria} (ME) phase, the distribution of $q$ shifts towards strictly positive values and broadens; simultaneously, the update typically become oscillatory or fail to converge, signaling the breakdown of the single-fixed-point assumption and the presence of a fragmented landscape. Finally, in the \emph{Unbounded Growth} (UG) phase, the distribution of the mean abundance $\mu$ diverges during the iterations, indicating the absence of finite fixed points and corresponding to the existence of unbounded trajectories in the microscopic dynamics.

    The complete stability phase diagram in the $(m, \sigma)$ plane is shown in \cref{fig:phasediagram}, illustrating the critical role of interaction symmetry $\gamma$. For moderate connectivity ($K=10$, left panel), the size of the SE phase is strongly dependent on the reciprocity of the interactions. As $\gamma$ increases from $-1$ (predator-prey) to $1$ (symmetric competition/mutualism), the SE region shrinks systematically, consistent with mean-field predictions for fully connected systems~\cite{gallaDynamicallyEvolvedCommunity2018}. However, unlike the fully connected case, the sparse network exhibits a ME phase even under fully antisymmetric interactions ($\gamma = -1$). In the infinite-connectivity limit, the $\gamma = -1$ system does not enter a ME phase, instead remaining in a single stable equilibrium up to the UG transition. By contrast, finite connectivity introduces localized feedback loops that destabilize coexistence once the mean interaction $m$ becomes sufficiently negative.

    Moreover, for intermediate symmetry values ($\gamma > -1$), the boundary between the SE and ME phases is not horizontal as in the FC case \cite{gallaDynamicallyEvolvedCommunity2018}; it bends markedly towards the $\sigma = 0$ axis, eventually intersecting at the topological critical point $m_c(K)$. This bending reveals a nontrivial synergy between topology and disorder: structural frustration amplifies the effect of interaction heterogeneity, driving the system into a fragmented, multi-equilibrium state at disorder strengths much weaker than those predicted by DMFT in the fully-connected case.

    This structural fragility is further amplified at low connectivity ($K=3$, right panel). In this regime, the Single Equilibrium (SE) region is markedly narrower than for $K=10$, underscoring the increased sensitivity of sparse networks to local competitive imbalances. The synergy between topology and disorder also manifests itself as a stronger curvature of the SEME boundary towards the $\sigma=0$ axis, further constraining the stability domain. This contraction reflects the breakdown of self-averaging within small neighborhoods: with only three neighbors, local fluctuations and asymmetric interaction motifs can disrupt coexistence more readily than in denser networks. Finally, for fully antisymmetric interactions ($\gamma=-1$), the ME phase shifts towards progressively more negative $m$ as connectivity $K$ increases, disappearing entirely in the limit $K \to \infty$. This behavior aligns seamlessly with the fully connected case, where no ME phase exists for $\gamma=-1$.

    Our work demonstrates that sparsity acts as a decisive organizing principle in ecosystem dynamics: finite connectivity does not merely weaken interactions but fundamentally restructures the stability landscape. By embedding the generalized Lotka-Volterra equations into a cavity process, we provided a complete characterization of the stability phase diagram in the thermodynamic limit. This analysis reveals that the topological phase transition to multiple equilibria--previously observed in homogeneous graphs--robustly persists in the presence of couplings heterogeneity. This structural mechanism is isolated most clearly in fully antisymmetric ($\gamma=-1$) predator-prey communities, where a fragmented attractor landscape emerges despite being strictly forbidden in the fully connected limit.
    
    By bridging the cavity method with ecological dynamics, we provide a scalable, analytically controlled framework that reconciles microscopic network structure with macroscopic resilience. The resulting phase diagram reveals how biodiversity and stability coevolve under structural constraints, offering new theoretical basis for understanding regime shifts, and context-dependent responses in real ecosystems. This approach establishes a new blueprint for studying non-equilibrium collective phenomena in sparse, adaptive systems--from neural circuits to microbial consortia--where topology and dynamics are inseparably intertwined. Indeed, the strength of the method lies in its universality. Unlike standard mean-field approximations, it does not rely on Gaussian interaction statistics or regular topologies (RRGs), but applies equally to systems with arbitrary coupling distributions and heterogeneous degree profiles. 

    M. T. thanks Ada Altieri, Fabián Aguirre-López, Joseph W. Baron for many stimulating discussions. M. T. and R. M. also thanks David Machado for his valuable comments on the work. The authors acknowledge financial support from the European REA, Marie Skłodowska-Curie Actions, grant agreement no. 101131463 (SIMBAD). Computational resources provided byht HPC@PoliTO (www.hpc.polito.it).

    %

\clearpage
\widetext
\begin{center}
\textbf{\large Supplemental material for ``Sparse Interactions Reshape Stability in Random Lotka–Volterra Dynamics''}
\end{center}

\setcounter{equation}{0}
\setcounter{figure}{0}
\setcounter{table}{0}
\setcounter{page}{1}
\setcounter{section}{0}

\renewcommand{\theequation}{S\arabic{equation}}
\renewcommand{\thefigure}{S\arabic{figure}}
\renewcommand{\thetable}{S\arabic{table}}

\renewcommand{\thesection}{S\Roman{section}} 

\setcounter{secnumdepth}{3} 

\renewcommand{\bibnumfmt}[1]{[S#1]}
\renewcommand{\citenumfont}[1]{S#1}


\section{\label{supp:introduction}Introduction}
    This Supplemental Material presents the detailed derivations and numerical procedures underlying the results of the main text. We begin by specifying the generalized Lotka–Volterra (gLV) model and the random-network ensemble considered in this work. The system consists of $N$ interacting species with abundances $x_i(t)$, whose interactions are encoded in a sparse, undirected graph with adjacency matrix $A=\{a_{ij}\}$. An entry $a_{ij}=a_{ji}=1$ denotes a direct interaction between species $i$ and $j$, while $a_{ij}=0$ otherwise. The network is sampled once from a quenched random ensemble with degree distribution $p_{\rm deg}(k)$; in the main text we focus on random regular graphs (RRGs), in which every species has exactly $K$ neighbors. This preserves sparsity while ensuring analytic tractability.

    The temporal evolution of species abundances is governed by the generalized Lotka–Volterra equations
    \begin{equation}
        \frac{d x_i(t)}{dt} = x_i(t) \left[1 - x_i(t) + \sum_{j\neq i} a_{ij} J_{ij} x_j(t)\right]. \label{eq:LV_eq_SM}
    \end{equation}
    For each interacting pair $(i,j)$, the coefficients $(J_{ij},J_{ji})$ are drawn from a quenched bivariate Gaussian ensemble with
    \[
        \overline{J_{ij}}=\frac{m}{K}, \qquad \operatorname{Var}(J_{ij})=\frac{\sigma^2}{K}, \qquad \operatorname{Corr}(J_{ij},J_{ji})=\gamma,
    \]
    where the overbar denotes disorder averaging. The $1/K$ scaling guarantees a proper thermodynamic limit as $K\to\infty$, recovering the fully connected case. The symmetry parameter $\gamma\in[-1,1]$ controls the structure of pairwise interactions: $\gamma=-1$ corresponds to fully antisymmetric (predator–prey–dominated) interactions, whereas $\gamma=1$ yields fully symmetric mutualistic or competitive pairs, with intermediate values interpolating smoothly between these ecological regimes.
    
    The remainder of this Supplemental Material is organized as follows. \Cref{supp:effective_cav_proc} derives the dynamic cavity equations appropriate for sparse interaction networks, starting from the Martin–Siggia–Rose–Janssen–De~Dominicis (MSRJD) functional formalism. Performing a small-coupling expansion of these equations we obtain the effective single-species stochastic process driven by self-consistent memory kernels and colored noise. \Cref{supp:fixed_point_equations} derives the corresponding fixed-point equations for the cavity means, variances, and susceptibilities. \Cref{supp:population_dynamics} describes the numerical implementation of the population-dynamics algorithm used to solve these equations, including update rules and convergence diagnostics. Finally, \cref{supp:phase_diagrams} explains how the phase boundaries reported in the main text were extracted from the behavior of the cavity fields.

\section{\label{supp:effective_cav_proc}Derivation of the effective cavity process}
    The calculation is based on the principles of \cite{taraboloGaussianApproximationDynamic20252}.
    The dynamical partition function of the Lotka-Volterra system can
    be written as
    \begin{equation}
        Z = \int\mathcal{D}[\underline{x}]\prod_{i}p_{0}\left(x_{i}\left(0\right)\right)\delta\left[\dot{x}_i(t) - x_i(t) \left(1 - x_i(t) + \sum_{j\neq i} a_{ij} J_{ij} x_j(t)\right)\right],
    \end{equation}
    where $\int\mathcal{D}[\underline{x}]=\prod_{i}\int\mathcal{D}[x]$ is a product of functional integrals, $\langle\cdot\rangle_{\underline{\eta}}=\left(\prod_{i}\int\mathcal{D}[\eta_{i}]\right)\cdot$ is the average over the noise realizations, and $\delta[\cdot]$ is a functional Dirac delta which restricts the integral to the paths that satisfy the stochastic dynamics. Using the Fourier representation of the delta functional $\delta[x]\propto\int\mathcal{D}[\hat{x}]\,e^{-{\rm i}\int dt\,\hat{x}(t)x(t)}$ we obtain
    \begin{equation}
        Z = \int\mathcal{D}[\underline{x},\underline{\hat{x}}]\prod_{i}p_{0}\left(x_{i}(0)\right)e^{-{\rm i}\int dt\,\hat{x}_{i}(t)\left[\dot{x}_{i}(t) - x_{i}(t)\left(1 - x_{i}(t) + \sum_{j\neq i} a_{ij} J_{ij} x_{j}(t)\right)\right]}.
    \end{equation}
    
    The integral over time runs over a fixed time interval, but we omit the limits of integration, here and elsewhere, for the sake of simplifying the notation. We introduce an auxiliary variable $f_i=1-x_i-\sum_{j\neq i}a_{ij}J_{ij}x_j(t)+h_i(t)$, which can be inserted in the calculation through a Dirac delta function. Integrating over the Gaussian noise we obtain
    \begin{align}
        Z = \int\mathcal{D}[\underline{x},\underline{\hat{x}},\underline{f},\underline{\hat{f}}]\prod_{i}p_{0}\left(x_{i}(0)\right)e^{S^{\rm loc}_i\left[x_i,\hat{x}_i,f_i,\hat{f}_i\right]}\prod_{(i,j)}e^{S^{\rm int}_{ij}\left[x_i,\hat{f}_i,x_j,\hat{f}_j\right]},
    \end{align}
    where we have defined the local and interactive actions as
    \begin{align}
      S^{\rm loc}_i & = -{\rm i}\int dt \left\{\hat{x}_i(t)\left(\dot{x}_i(t)-x_i(t)f_i(t)\right)+\hat{f}_i(t)\left(f_i(t)-1+x_i(t)\right) \right\},\\
      S^{\rm int}_{ij} & = {\rm i}\int dt\left(J_{ij}\hat{f}_i(t)x_j(t)+J_{ji}\hat{f}_j(t)x_i(t) \right).
    \end{align}
    
    The dynamical partition function appears to be factorized according to a graphical model with the same structure as the underlying interaction graph.  According to this graphical model construction, we propose a dynamic cavity ansatz to approximate marginal distributions of abundances of a rGLV on sparse networks. This yields a set of fixed point equations for the cavity messages
    \begin{equation}
      c_{i\setminus j}[x_{i},\hat{x}_{i}, f_i, \hat{f}_i] = \frac{p_{0}\left(x_i(0)\right)}{Z_{i\setminus j}}e^{S^{\rm loc}_i}\prod_{k\in\partial i \setminus j}\int \mathcal{D}[x_k,\hat{x}_k,f_k,\hat{f}_k] \, c_{k\setminus i}[x_{k},\hat{x}_{k}, f_k, \hat{f}_k] e^{S^{\rm int}_{ki}},\label{eq:cavity_marg}
    \end{equation}
    where $\partial i\setminus j$ is the set of neighboring indices of $i$ except for $j$. The cavity equations should be interpreted as a message-passing procedure, where messages are iteratively exchanged between nodes until convergence to a fixed point.
    
    Finally, completing the cavity and computing the total marginal over $i$ gives
    \begin{equation}
      c_{i}[x_{i},\hat{x}_{i}, f_i, \hat{f}_i]=\frac{p_{0}\left(x_i(0)\right)}{Z_{i}}e^{S^{\rm loc}_i}\prod_{k\in\partial i}\int \mathcal{D}[x_k,\hat{x}_k,f_k,\hat{f}_k] \, c_{k\setminus i}[x_{k},\hat{x}_{k}, f_k, \hat{f}_k] e^{S^{\rm int}_{ki}}.\label{eq:marg_RRG}
    \end{equation}
    
    The normalization $Z_{i\setminus j}$ of the messages at each step is not critical and can be adjusted after convergence. However, we choose to properly normalize the messages by defining the cavity normalization factor as
    \begin{equation}
        Z_{i\setminus j} = \int \mathcal{D}[x_i,\hat{x}_i,f_i,\hat{f}_i] \, p_{0}\left(x_i(0)\right)e^{S^{\rm loc}_i}\prod_{k\in\partial i\setminus j}\int \mathcal{D}[x_k,\hat{x}_k,f_k,\hat{f}_k] \, c_{k\setminus i}[x_{k},\hat{x}_{k}, f_k, \hat{f}_k] e^{S^{\rm int}_{ki}}.
    \end{equation}
    This ensures that the cavity messages are properly normalized quasi-probability distributions and allows for the definition of averages over them. Consequently, the expectation of any generic function $g(x_i, \hat{x_i}, f_i, \hat{f}_i)$ over the cavity distribution can be computed as
    \begin{equation}
        \left\langle g(x_i, \hat{x_i}, f_i, \hat{f}_i) \right\rangle_{i\setminus j} = \int \mathcal{D}[x_i,\hat{x}_i,f_i,\hat{f}_i] \, c_{i\setminus j}[x_{i},\hat{x}_{i}, f_i, \hat{f}_i] g(x_i, \hat{x_i}, f_i, \hat{f}_i).
    \end{equation}
    
    We perform a second-order expansion in the coupling parameters $J_{ij}$ of the cavity messages defined in \cref{eq:cavity_marg}, following the approach of \cite{taraboloGaussianApproximationDynamic20252}. This yields:
    \begin{align}
        c_{i\setminus j}[x_i, \hat{x}_i, f_i, \hat{f}_i] & \approx \frac{p_0(x_i(0))}{Z_{i\setminus j}} e^{S^{\mathrm{loc}}_i} \prod_{k \in \partial i \setminus j} \left\langle 1 + S^{\mathrm{int}}_{ki} + \frac{1}{2}(S^{\mathrm{int}}_{ki})^2 + \dots \right\rangle_{k\setminus i} \\
        &= \frac{p_0(x_i(0))}{Z_{i\setminus j}} e^{S^{\mathrm{loc}}_i} \prod_{k \in \partial i \setminus j} \Bigg[ 1 
        + \mathrm{i} J_{ik} \int dt \, \hat{f}_i(t) \langle x_k(t) \rangle_{k\setminus i} + \frac{J_{ik} J_{ki}}{2} \, \mathrm{i} \int dt \, dt' \, \hat{f}_i(t) \langle x_k(t) \, \mathrm{i} \hat{f}_k(t') \rangle_{k\setminus i} x_i(t') \nonumber \\
        &\quad \left. + \frac{J_{ik}^2}{2} \int dt \, dt' \, {\rm i}\hat{f}_i(t) \langle x_k(t) x_k(t') \rangle_{k\setminus i} {\rm i}\hat{f}_i(t') \right],
    \end{align}
    where we neglect averages involving only hatted quantities, since these vanish identically \cite{taraboloGaussianApproximationDynamic20252}. Re-exponentiating the terms inside the product yields an approximate expression for the cavity marginal in terms of an effective single-site action:
    \begin{equation}
        c_{i\setminus j}[x_i, \hat{x}_i, f_i, \hat{f}_i] \approx \frac{p_0(x_i(0))}{Z_{i\setminus j}} e^{S^{\mathrm{eff}}_{i\setminus j}},
    \end{equation}
    with the effective action given by
    \begin{align}
        S^{\mathrm{eff}}_{i\setminus j} &= -\mathrm{i} \int dt \Bigg\{ \hat{x}_i(t) \left[\dot{x}_i(t) - x_i(t) f_i(t)\right] 
        + \hat{f}_i(t) \Big[ f_i(t) - 1 + x_i(t)  - \sum_{k \in \partial i \setminus j} J_{ik} \mu_{k\setminus i}(t) \nonumber \\
        &\quad 
        - \sum_{k \in \partial i \setminus j} J_{ik} J_{ki} \int dt' \, R_{k\setminus i}(t,t') x_i(t') \Big] \Bigg\} + \int dt \, dt' \, \mathrm{i} \hat{f}_i(t) 
        \left( \sum_{k \in \partial i \setminus j} J_{ik}^2 \, C_{k\setminus i}(t,t') \right) \mathrm{i} \hat{f}_i(t'),
    \end{align}
    where we have defined the cavity means, response functions, and correlations as
    \begin{align}
        \mu_{k\setminus i}(t) &= \langle x_k(t) \rangle_{k\setminus i}, \\
        R_{k\setminus i}(t,t') &= \langle x_k(t) \, \mathrm{i} \hat{f}_k(t') \rangle_{k\setminus i}, \\
        C_{k\setminus i}(t,t') &= \langle x_k(t) x_k(t') \rangle_{k\setminus i} - \mu_{k\setminus i}(t) \mu_{k\setminus i}(t').
    \end{align}
    
    The effective action can be interpreted as the MSRJD action of an effective Langevin equation. Integrating over the auxiliary fields $f_i$ and $\hat{f}_i$ yields the effective cavity process for species $i$ in the absence of neighbor $j$:
    \begin{equation}
        \frac{d x_{i\setminus j}(t)}{dt} = x_{i\setminus j}(t) \left[ 1 - x_{i\setminus j}(t) + \sum_{k \in \partial i \setminus j} J_{ik} \mu_{k\setminus i}(t) + \sum_{k \in \partial i \setminus j} J_{ik} J_{ki} \int_0^t dt' \, R_{k\setminus i}(t,t') x_{i\setminus j}(t') + h_{i\setminus j}(t) + \xi_{i\setminus j}(t) \right], \label{eq:eff_proc_cav2}
    \end{equation}
    where $h_{i\setminus j}(t)$ is an external field, needed to compute response functions, and $\xi_{i\setminus j}$ is a Gaussian noise with vanishing mean and correlation 
    \begin{equation}
        \langle \xi_{i\setminus j}(t) \xi_{i\setminus j}(t') \rangle_{i\setminus j} = \sum_{k \in \partial i \setminus j} J_{ik}^2 \, C_{k\setminus i}(t,t').
    \end{equation}
    The terms $\mu_{k\setminus i}(t)$, $R_{k\setminus i}(t,t')$, and $C_{k\setminus i}(t,t')$ are respectively the cavity mean abundance, the cavity response function, and the cavity (connected) two-point correlation function. They are obtained self-consistently from the effective dynamics of the neighbors
    \begin{align}
        \mu_{k\setminus i}(t)&=\left\langle x_{k\setminus i}(t)\right\rangle_{k\setminus i},\\
        R_{k\setminus i}(t,t')&=\frac{\delta}{\delta h_{k\setminus i}(t')}\left\langle x_{k\setminus i}(t)\right\rangle _{k\setminus i},\\
        C_{k\setminus i}(t,t')&=\left\langle x_{k\setminus i}(t)x_{k\setminus i}(t')\right\rangle _{k\setminus i}-\mu_{k\setminus i}(t)\mu_{k\setminus i}(t'),
    \end{align}
    where $\left\langle \cdot\right\rangle _{k\setminus i}$ denotes the average over the distribution of the effective process \cref{eq:eff_proc_cav2} for the edge $(k,i)$. The effective process is therefore de-coupled, and the interactions are captured by the self-consistent external fields $\mu_{k\setminus i}(t)$, $R_{k\setminus i}(t,t')$, and $C_{k\setminus i}(t,t')$.
    
    Once the cavity averages have been determined through self-consistency, one can obtain the effective de-coupled Langevin equations for the species population
    \begin{equation}
        \frac{d x_{i}(t)}{dt} = x_{i}(t)\Biggl[1-x_{i}(t)+\sum_{k\in\partial i}J_{ik}\mu_{k\setminus i}(t) + \sum_{k\in\partial i}\int_{0}^{t}dt'J_{ik}J_{ki}R_{k\setminus i}(t,t')x_{i}(t') + h_{i}(t)+\xi_{i}(t)\Biggr],\label{eq:eff_proc_marg}
    \end{equation}
    with Gaussian noise with vanishing mean and correlation $\left\langle \xi_{i}(t)\xi_{i}(t')\right\rangle _{i}=\sum_{k\in\partial i}J_{ik}^{2}C_{k\setminus i}(t,t')$.
    The average $\left\langle \cdot\right\rangle _{i}$ is performed over the effective de-coupled process \cref{eq:eff_proc_marg}.
    
\section{\label{supp:fixed_point_equations}Fixed-point equations}
    To proceed analytically, we assume that the system relaxes to a fixed point independent of the initial conditions, i.e.
    \begin{equation}
        \lim_{t\to\infty} x_i(t) = x_i^{\star}.
    \end{equation}
    \begin{equation}
            \lim_{t \to \infty} x_{i\setminus j}(t) = x_{i\setminus j}^{\star}, \qquad 
            \lim_{t \to \infty} \xi_{i\setminus j}(t) = \xi_{i\setminus j}^{\star}.
    \end{equation} 
    Consequently, the mean cavity abundance and the two-point correlation function become stationary,
    \begin{align}
        \lim_{t \to \infty} \mu_{i\setminus j}(t) &= \mu_{i\setminus j}^{\star} = \langle x_{i\setminus j}^{\star} \rangle_{i\setminus j}^{\star},\\
        \lim_{t, t' \to \infty} C_{i\setminus j}(t,t') &= q_{i\setminus j}^{\star} = \langle (x_{i\setminus j}^{\star})^2 \rangle_{i\setminus j}^{\star} - (\mu_{i\setminus j}^{\star})^2 ,
    \end{align}
    where $\langle \cdot \rangle_{i\setminus j}^{\star}$ denotes the average over the stationary cavity process. 
    Under this assumption, the cavity response function becomes time-translationally invariant and depends only on the time difference $\tau = t - t'$. 
    We define its time integral,
    \begin{equation}
        \chi_{i\setminus j}^{\star} = \int_0^{\infty} d\tau\, R_{i\setminus j}^{\star}(\tau),
    \end{equation}
    which represents the integrated response (or susceptibility) of the cavity process.
    
     At the fixed point, both $x_{i\setminus j}^{\star}$ and the effective field $\xi_{i\setminus j}^{\star}$ become static random variables. 
    The variable $x_{i\setminus j}^{\star}$ depends functionally on $\xi_{i\setminus j}^{\star}$, which is Gaussian distributed with zero mean and variance
    \begin{equation}
        \langle (\xi_{i\setminus j}^{\star})^2 \rangle_{i\setminus j}^{\star}
        = \sum_{k \in \partial i \setminus j} J_{ik}^2\, q_{k\setminus i}^{\star}
        = \Sigma_{i\setminus j}^2.
    \end{equation}
    
    Since $x_{i\setminus j}^\star$ is constant in time, the fixed point abundances satisfy \cref{eq:eff_proc_cav2} with vanishing time derivative, i.e.
    \begin{align}
        0 &= x_{i\setminus j}^{\star} \bigg[ 1 - x_{i\setminus j}^{\star} + \sum_{k\in\partial i\setminus j} J_{ik}\mu_{k\setminus i}^{\star} + \sum_{k\in\partial i\setminus j} J_{ik}J_{ki}\chi_{k\setminus i}^{\star}x_{i\setminus j}^{\star} + \xi_{i\setminus j}^{\star} + h_{i\setminus j} \bigg].
    \end{align}
    These equations admit two possible solutions: one corresponding to extinction, $x_{i\setminus j}^\star=0$, and the other to a non-vanishing equilibrium abundance,
    \begin{equation}
        x_{i\setminus j}^\star = \frac{1 + \sum_{k\in \partial i \setminus j} J_{ik} \mu_{k\setminus i}^\star + \xi_{i \setminus j}^\star + h_{i \setminus j}}{1 - \sum_{k\in \partial i \setminus j} J_{ik} J_{ki} \chi_{k \setminus i}^\star},
    \end{equation}
    which is physically admissible only if the right-hand side is positive. 
    The fixed-point solution can thus be written as
    \begin{equation}
        x_{i\setminus j}^\star(\xi_{i\setminus j}^\star) = \frac{1 + \sum_{k\in \partial i \setminus j} J_{ik} \mu_{k\setminus i}^\star + \xi_{i \setminus j}^\star + h_{i \setminus j}}{1 - \sum_{k\in \partial i \setminus j} J_{ik} J_{ki} \chi_{k \setminus i}^\star}\; \Theta\!\left( \frac{1 + \sum_{k\in \partial i \setminus j} J_{ik} \mu_{k\setminus i}^\star + \xi_{i \setminus j}^\star + h_{i \setminus j}}{1 - \sum_{k\in \partial i \setminus j} J_{ik} J_{ki} \chi_{k \setminus i}^\star} \right).
    \end{equation}
    
    We now introduce the short-hand notations
    \begin{align}
        \Delta_{i\setminus j} &= \frac{1 + \sum_{k\in\partial i\setminus j} J_{ik}\mu_{k\setminus i}^{\star}}{\Sigma_{i\setminus j}},\\
        \Gamma_{i\setminus j} &= 1 - \sum_{k\in\partial i\setminus j} J_{ik}J_{ki}\chi_{k\setminus i}^{\star}.
    \end{align}
    Since $\xi_{i\setminus j}^{\star}$ is Gaussian with mean zero and variance $\Sigma_{i\setminus j}^2$, we set $\xi_{i\setminus j}^{\star} = -\Sigma_{i\setminus j}\, s$, where $s$ is a standard normal variable with zero mean and unit variance. 
    In this way, the fixed point can be expressed as a function of $s$ only:
    \begin{align}
        x_{i\setminus j}^{\star}(s)
        &= \frac{\Sigma_{i\setminus j}(\Delta_{i\setminus j}-s) + h_{i\setminus j}}{\Gamma_{i\setminus j}}
        \, \Theta\!\left(\frac{\Sigma_{i\setminus j}(\Delta_{i\setminus j}-s) + h_{i\setminus j}}{\Gamma_{i\setminus j}}\right).
    \end{align}
    
    The random variable $x_{i\setminus j}^{\star}$ has moments $\mu_{i\setminus j}^{\star}$, $q_{i\setminus j}^{\star}$, and susceptibility $\chi_{i\setminus j}^{\star}$, which are obtained self-consistently by averaging over the Gaussian variable $s$. 
    The mean reads
    \begin{align}
        \mu_{i\setminus j}^\star
        &= \left . \langle x_{i\setminus j}^{\star} \rangle_{i\setminus j}^\star \right|_{h_{i\setminus j}=0}
         =  \left . \int \frac{ds\, e^{-s^2/2}}{\sqrt{2 \pi}}\, x_{i\setminus j}^\star(s) \right|_{h_{i\setminus j}=0} \nonumber\\
        &= \frac{\Sigma_{i\setminus j}}{\Gamma_{i\setminus j}}\left[\Theta(\Gamma_{i\setminus j})\int_{-\infty}^{\Delta_{i\setminus j}}\!Ds\,(\Delta_{i\setminus j}-s)
        + \Theta(-\Gamma_{i\setminus j})\int_{\Delta_{i\setminus j}}^{\infty}\!Ds\,(\Delta_{i\setminus j}-s)\right]\label{eqapp:mucav_fixedpoint},
    \end{align}
    while the variance is
    \begin{align}
        q_{i\setminus j}^\star
        &=  \left . \langle (x_{i\setminus j}^{\star})^2 \rangle_{i\setminus j}^\star  \right|_{h_{i\setminus j}=0}
          - (\mu_{i\setminus j}^\star)^2
         =   \left . \int \frac{ds\, e^{-s^2/2}}{\sqrt{2 \pi}}\, \big(x_{i\setminus j}^\star(s)\big)^2 \right|_{h_{i\setminus j}=0}  - (\mu_{i\setminus j}^\star)^2 \nonumber\\
        &= \frac{\Sigma^2_{i\setminus j}}{\Gamma^2_{i\setminus j}}\left[\Theta(\Gamma_{i\setminus j})\int_{-\infty}^{\Delta_{i\setminus j}}\!Ds\,(\Delta_{i\setminus j}-s)^2
        + \Theta(-\Gamma_{i\setminus j})\int_{\Delta_{i\setminus j}}^{\infty}\!Ds\,(\Delta_{i\setminus j}-s)^2\right]  - (\mu_{i\setminus j}^\star)^2\label{eqapp:qcav_fixedpoint},
    \end{align}
    where $Ds = ds\, e^{-s^2/2} / \sqrt{2\pi}$. 
    The susceptibility is obtained by differentiating the average fixed-point abundance with respect to a constant external field $h_{i\setminus j}$:
    \begin{align}
        \chi_{i\setminus j}^\star
        &= \left . \frac{\partial \langle x_{i\setminus j}^{\star} \rangle_{i\setminus j}^\star}{\partial h_{i\setminus j}} \right|_{h_{i\setminus j}=0}
         = \frac{1}{\Gamma_{i\setminus j}}\left[\Theta(\Gamma_{i\setminus j})\int_{-\infty}^{\Delta_{i\setminus j}}\!Ds \;
         + \Theta(-\Gamma_{i\setminus j})\int_{\Delta_{i\setminus j}}^{\infty}\!Ds\right]\label{eqapp:chicav_fixedpoint}.
    \end{align}
    
    We observe that these integrals can be written in terms of the error function $\operatorname{erf}(x)=\int_0^{x} ds\,e^{-s^2}2/\sqrt{\pi}$. 
    Defining $\varphi(x)=e^{-x^2/2}/\sqrt{2\pi}$ as the standard normal density and $\Phi(x)=[1+\operatorname{erf}(x/\sqrt{2})]/2$ its cumulative distribution function, the integrals can be expressed as moments of truncated normal distributions. 
    In particular,
    \begin{align*}
        \int_{-\infty}^{\Delta} Ds        &= \Phi(\Delta),                           &     \int_{\Delta}^{\infty} Ds        &= \Phi(-\Delta),                              \\
        \int_{-\infty}^{\Delta} Ds \, s   &= \varphi(\Delta),                        &     \int_{\Delta}^{\infty} Ds \, s   &= \varphi(\Delta),                            \\
        \int_{-\infty}^{\Delta} Ds \, s^2 &= \Phi(\Delta) - \Delta \varphi(\Delta),  &     \int_{\Delta}^{\infty} Ds \, s^2 &= \Phi(-\Delta) + \Delta \varphi(\Delta),
    \end{align*}
    from which, substituting into \cref{eqapp:mucav_fixedpoint,eqapp:qcav_fixedpoint,eqapp:chicav_fixedpoint}, we obtain the cavity update equations
    \begin{align}
        \mu_{i\setminus j}^{\star} &= \mathcal{F}_{\mu}\!\left(\Delta_{i\setminus j}, \Sigma_{i\setminus j}, \Gamma_{i\setminus j}\right)\label{eq:mucav_fixedpoint},\\
        q_{i\setminus j}^{\star} &= \mathcal{F}_{q}\!\left(\Delta_{i\setminus j}, \Sigma_{i\setminus j}, \Gamma_{i\setminus j}\right) - (\mu_{i\setminus j}^{\star})^2\label{eq:qcav_fixedpoint},\\
        \chi_{i\setminus j}^{\star} &= \mathcal{F}_{\chi}\!\left(\Delta_{i\setminus j}, \Gamma_{i\setminus j}\right)\label{eq:chicav_fixedpoint},
    \end{align}
    where the update functions are given explicitly by
    \begin{align}
        \mathcal{F}_{\mu}\!(\Delta, \Sigma, \Gamma) &= \frac{\Sigma}{\Gamma} \bigg[ \Theta(\Gamma) \Big(\Delta \Phi(\Delta) + \varphi(\Delta) \Big) + \Theta(-\Gamma) \Big(\Delta\Phi(-\Delta) - \varphi(-\Delta)\Big) \bigg] \label{eq:fmu_update}\\
        \mathcal{F}_{q}(\Delta, \Sigma, \Gamma) &= \frac{\Sigma^2}{\Gamma^2} \bigg[ \Theta(\Gamma) \Big((1+\Delta^2)\Phi(\Delta) + \Delta\varphi(\Delta)\Big) + \Theta(-\Gamma) \Big((1+\Delta^2)\Phi(-\Delta) - \Delta\varphi(-\Delta)\Big) \bigg], \label{eq:fq_update}\\
        \mathcal{F}_{\chi}\!(\Delta, \Gamma) &= \frac{1}{\Gamma} \big[ \Theta(\Gamma)\,\Phi(\Delta) + \Theta(-\Gamma)\,\Phi(-\Delta) \big]. \label{eq:fchi_update}
    \end{align}

    An equivalent set of equations holds for the full marginals, describing the species abundances in the original graph. Denoting the fixed point moments by $\mu_i^{\star}$, $q_i^{\star}$, and $\chi_i^{\star}$, we have
    \begin{align}
        \mu_i^{\star} &= \mathcal{F}_{\mu}\!\left(\Delta_i, \Sigma_i, \Gamma_i\right),\label{eq:mufull_fixedpoint}\\
        q_i^{\star} &= \mathcal{F}_{q}\!\left(\Delta_i, \Sigma_i, \Gamma_i\right) - (\mu_i^{\star})^2,\label{eq:qfull_fixedpoint}\\
        \chi_i^{\star} &= \mathcal{F}_{\chi}\!\left(\Delta_i, \Gamma_i\right),\label{eq:chifull_fixedpoint}
    \end{align}
    where the update functions are the same as in \cref{eq:fmu_update,eq:fq_update,eq:fchi_update}, and we have defined
    \begin{align}
        \Delta_i &= \frac{1 + \sum_{k\in\partial i} J_{ik}\mu_{k\setminus i}^{\star}}{\Sigma_i},\label{eq:Delta_full_def}\\
        \Sigma_i^2 &= \sum_{k \in \partial i} J_{ik}^2\, q_{k\setminus i}^{\star},\label{eq:Sigma_full_def}\\
        \Gamma_i &= 1 - \sum_{k\in\partial i} J_{ik}J_{ki}\chi_{k\setminus i}^{\star}.\label{eq:Gamma_full_def}
    \end{align}

    The coupled set of nonlinear equations \cref{eq:mucav_fixedpoint,eq:qcav_fixedpoint,eq:chicav_fixedpoint} defines the fixed point statistics of the cavity processes in terms of their local fields. These relations can be solved numerically via a message-passing procedure that iteratively updates the cavity quantities by substituting the moments of neighboring species.
        
\section{\label{supp:population_dynamics}Population dynamics algorithm}
    To compute disorder-averaged observables in the thermodynamic limit, we solve the self-consistent equations \cref{eq:mucav_fixedpoint,eq:qcav_fixedpoint,eq:chicav_fixedpoint} through a population dynamics algorithm. This stochastic iterative method propagates the distribution of the cavity fields under the joint randomness of network topology and interaction disorder, providing an efficient way to sample the typical stationary state of large random ecosystems.

    The algorithm proceeds as follows:
    \begin{enumerate}
        \item Initialize a population $\mathcal{P}$ of $P$ cavity variables, each represented by a triplet $(\mu^{\rm c}, q^{\rm c}, \chi^{\rm c})$ corresponding to the mean abundance, variance, and susceptibility of a cavity process.
        \item At each iteration:
        \begin{enumerate}
            \item Sample a degree $k$ from the cavity degree distribution
            \begin{equation*}
                p_{\rm deg}^{\rm c}(k) = \frac{k+1}{K}\, p_{\rm deg}(k+1).
            \end{equation*}
            \item randomly-draw $k$ triplets
            $\{(\mu_j^{\rm c}, q_j^{\rm c}, \chi_j^{\rm c})\}_{j=1}^{k}$ from the current population to represent the neighbors in the cavity graph.
            \item Draw $k$ pairs of random couplings $\{(J_j, J_j')\}_{j=1}^{k}$ from the coupling distribution.
            \item Compute the effective parameters:
            \begin{align*}
                \Delta^{\rm c} &= \frac{1 + \sum_{j=1}^{k} J_j \mu_j^{\rm c}}{\Sigma^{\rm c}},\\
                \Sigma^{\rm c} &= \sqrt{\sum_{j=1}^{k} J_j^2 q_j^{\rm c}},\\
                \Gamma^{\rm c} &= 1 - \sum_{j=1}^{k} J_j J_j' \chi_j^{\rm c}.
            \end{align*}
            \item Update the new fields according to
            \begin{align*}
                \mu^{\rm c} &= \mathcal{F}_{\mu}(\Delta^{\rm c}, \Sigma^{\rm c}, \Gamma^{\rm c}), \\
                q^{\rm c}   &= \mathcal{F}_{q}(\Delta^{\rm c}, \Sigma^{\rm c}, \Gamma^{\rm c}, \mu^{\rm c}), \\
                \chi^{\rm c}&= \mathcal{F}_{\chi}(\Delta^{\rm c}, \Gamma^{\rm c}).
            \end{align*}
            \item Replace softly a randomly-chosen element of the population with the new triplet $(\mu^{\rm c}, q^{\rm c}, \chi^{\rm c})$.
        \end{enumerate}
        \item Repeat until convergence of the population statistics, as assessed from the empirical averages of $\mu^{\rm c}$, $q^{\rm c}$, and $\chi^{\rm c}$.
    \end{enumerate}

    After convergence, single-site observables for the full graph can be estimated by performing one additional update step using degrees drawn from $p_{\rm deg}$ instead of $p_{\rm deg}^{\rm c}$, thus completing the cavity construction.

    \subsection{\label{supp:popdyn_details}Implementation details for the Population Dynamics algorithm}
        The population dynamics algorithm employed throughout this chapter is implemented in the open-source Julia package \texttt{RandomLotkaVolterraCavity.jl}~\cite{repo_rGLV2}. 
        Its structure follows the procedure outlined in \cref{supp:population_dynamics}, with additional numerical optimizations introduced to improve stability and convergence.
        
        A key modification concerns the way population elements are updated at each iteration. 
        Instead of replacing a single randomly-chosen element after every update, as in the standard population dynamics scheme, we proceed by generating a random permutation of the population indices at the beginning of each iteration (or \emph{sweep}). 
        The algorithm then sequentially updates each element according to this shuffled order, ensuring that every population member is updated exactly once per sweep while minimizing correlations arising from a fixed update sequence. 
        For each element in the shuffled population, the triplet $(\mu^{\mathrm{c}}, q^{\mathrm{c}}, \chi^{\mathrm{c}})$ is updated according to the self-consistent functions $\mathcal{F}_{\mu}$, $\mathcal{F}_{q}$, and $\mathcal{F}_{\chi}$ defined in \cref{eq:fmu_update,eq:fq_update,eq:fchi_update}.
        
        The update is performed softly by introducing a damping factor $d \in (0,1)$, which mixes the newly computed values with the previous ones:
        \begin{align}
            \mu^{\mathrm{c}} &\leftarrow d\, \mu^{\mathrm{c},\text{new}} + (1-d)\, \mu^{\mathrm{c},\text{old}}, \\
            q^{\mathrm{c}}   &\leftarrow d\, q^{\mathrm{c},\text{new}} + (1-d)\, q^{\mathrm{c},\text{old}}, \\
            \chi^{\mathrm{c}}&\leftarrow d\, \chi^{\mathrm{c},\text{new}} + (1-d)\, \chi^{\mathrm{c},\text{old}}.
        \end{align}
        This under-relaxation procedure suppresses oscillations and improves numerical stability, although it typically slows down the convergence rate. 
        In all the analyses presented in this work, the damping factor was chosen in the range $d \in [0.2, 0.6]$, depending on the convergence properties of the system.
        
        After completing a full sweep over the shuffled population, convergence is checked by comparing the average values of $\mu$, $q$, and $\chi$ before and after the sweep. 
        If the relative variation of these averages falls below a fixed threshold, the algorithm is stopped and convergence is assumed. 
        In our simulations, the tolerance threshold was set between $10^{-6}$ and $10^{-8}$, depending on the parameter regime.
        
        In principle, one should also verify the convergence of higher-order cumulants to ensure that the full population distribution has reached stationarity. 
        However, this requirement substantially increases computational time. 
        In practice, we found that monitoring the first moments provides reliable convergence for all parameter ranges explored in this chapter.
    
\section{\label{supp:phase_diagrams}Phase diagrams for random regular graphs}
    This section details the numerical procedures employed to construct the stability phase diagrams presented in the main text. We first describe the algorithm used to identify critical points from population dynamics data, and then present the phase diagrams for the specific values of connectivity $K$ and interaction symmetry $\gamma$ discussed in the main text, explaining how the critical boundaries were derived via interpolation of the numerical results.

    \subsection{\label{supp:critical_points}Numerical determination of critical points}
        The phase boundaries were determined by performing systematic parameter sweeps. For the topological transition ($\sigma=0$), we fixed the connectivity $K$ and varied the interaction mean $m$. For the general phase diagrams ($\sigma > 0$), we fixed $\sigma$ and scanned $m$ around the critical region. For each parameter set, we computed the distribution of the stationary moments ($\mu, q, \chi$) with the PopDyn algorithm, saving their average and standard deviation.
        
\begin{figure}[t!]
            \centering
            \includegraphics[width=.7\linewidth]{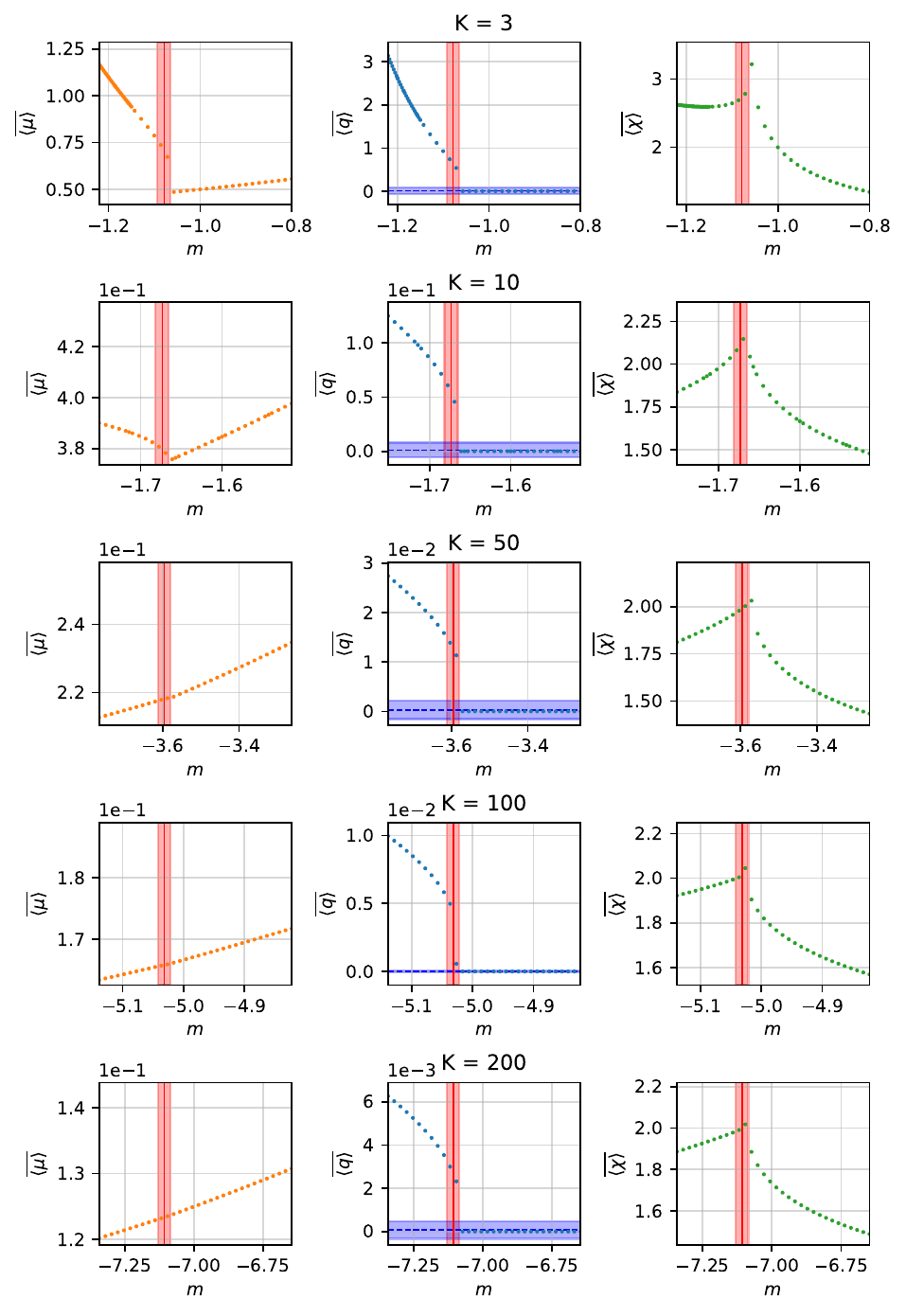}
            \caption{\label{fig:critical_example_topo} \textbf{Numerical determination of the topological phase transition ($\sigma=0$).} Each row corresponds to a different connectivity $K$. The columns display the population averages of the mean abundance $\overline{\langle\mu\rangle}$ (left), the cavity variance $\overline{\langle q\rangle}$ (center), and the susceptibility $\overline{\langle\chi\rangle}$ (right). The errors represent the standard deviation of the population distribution. In the central column ($q$), the blue horizontal line indicates the baseline noise floor (mean $\pm$ std), while the vertical red line marks the detected critical point $m_c$ (with its estimated error width). Note the sharp discontinuity in $q$ characteristic of the topological transition.}
        \end{figure}

\begin{figure}[t!]
            \centering
            \includegraphics[width=.7\linewidth]{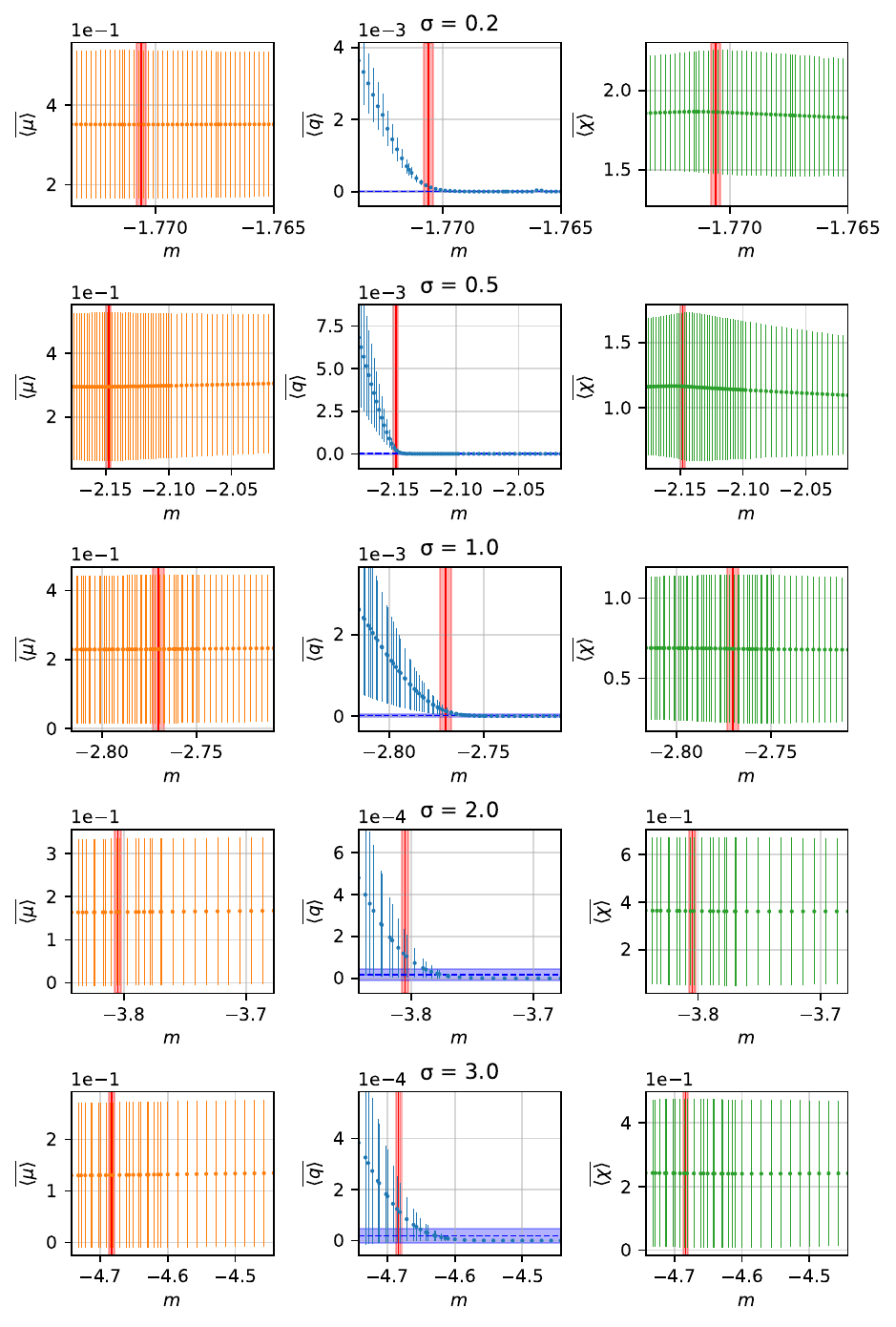}
            \caption{\label{fig:critical_example_SE_ME} \textbf{Detection of the Single Equilibrium (SE) to Multiple Equilibria (ME) transition.} Parameters: $K=10, \gamma=-1$. Each row corresponds to a different disorder strength $\sigma$. The columns display the population averages of the mean abundance $\overline{\langle\mu\rangle}$ (left), the cavity variance $\overline{\langle q\rangle}$ (center), and the susceptibility $\overline{\langle\chi\rangle}$ (right). The errors represent the standard deviation of the population distribution. In the central column ($q$), the blue horizontal line indicates the baseline noise floor (mean $\pm$ std), while the vertical red line marks the detected critical point $m_c$ (with its estimated error width). The transition is identified by the departure of the cavity variance $q$ from zero, while the mean abundance $\mu$ remains finite. Unlike the topological case, the order parameter grows continuously from zero, resulting in larger uncertainties for $m_c$ as the distribution slowly broadens.}
        \end{figure}

\begin{figure}[t!]
            \centering
            \includegraphics[width=.7\linewidth]{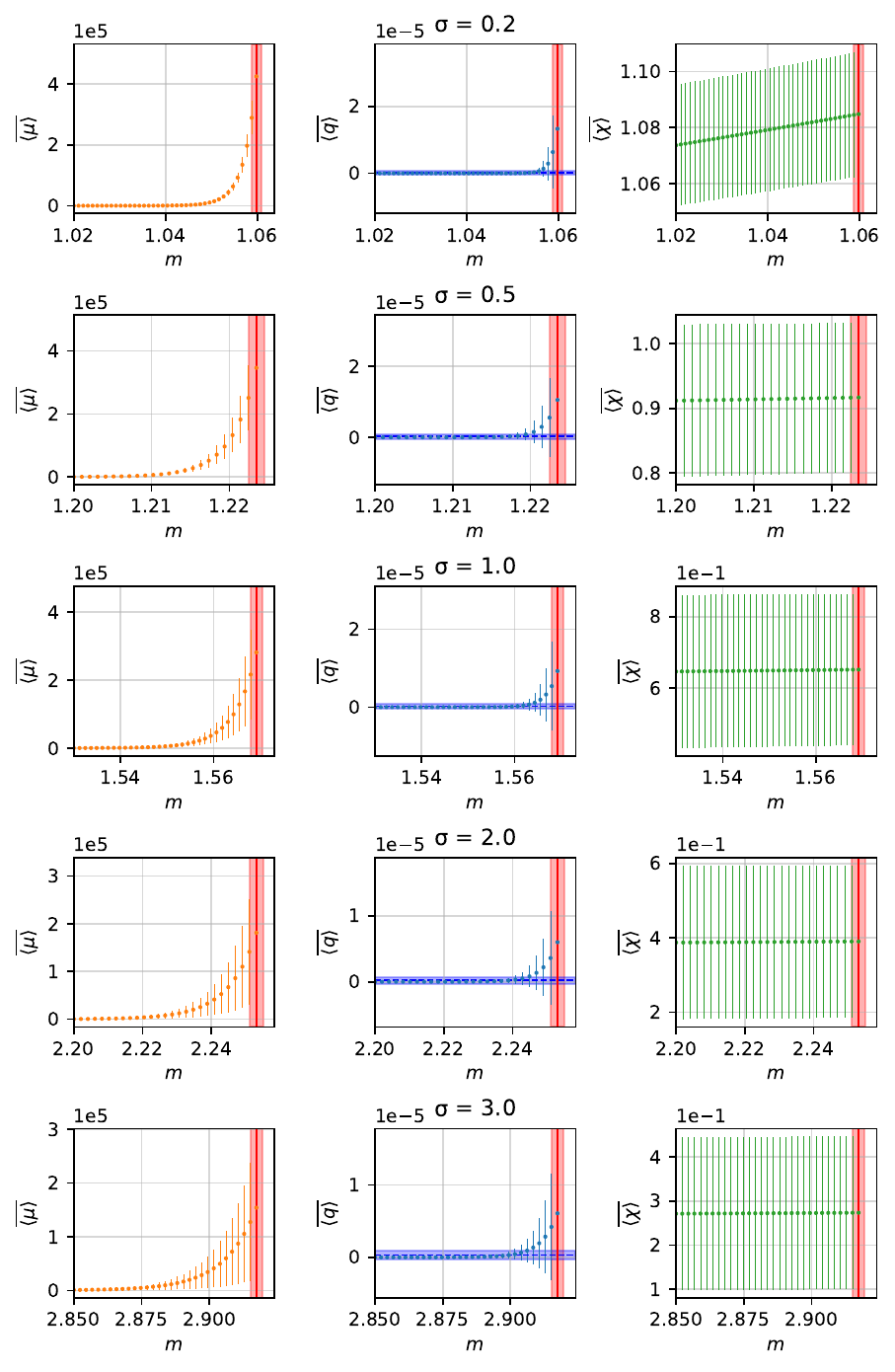}
            \caption{\label{fig:critical_example_SE_UG} \textbf{Detection of the Single Equilibrium (SE) to Unbounded Growth (UG) transition.} Parameters: $K=10, \gamma=-1$. Each row corresponds to a different disorder strength $\sigma$. The columns display the population averages of the mean abundance $\overline{\langle\mu\rangle}$ (left), the cavity variance $\overline{\langle q\rangle}$ (center), and the susceptibility $\overline{\langle\chi\rangle}$ (right). The errors represent the standard deviation of the population distribution. In the central column ($q$), the blue horizontal line indicates the baseline noise floor (mean $\pm$ std), while the vertical red line marks the detected critical point $m_c$ (with its estimated error width). The critical point is associated with the divergence of the mean abundance $\mu$. The vertical red line marks where the automated algorithm detects the instability, signaling the loss of linear stability and the onset of unbounded population growth.}
        \end{figure}

        To identify the critical transition points $m_c$, we implemented an automated algorithm designed to separate the signal from the numerical noise floor. The procedure is illustrated in \cref{fig:critical_example_topo,fig:critical_example_SE_ME,fig:critical_example_SE_UG}, and consists of the following steps:
        
        \begin{enumerate}
            \item \textbf{Baseline Characterization:} We define a baseline region using the first points of the scanned $m$-values, assumed to lie deep within the stable phase where the order parameter (e.g., the cavity variance $q$) should theoretically vanish. We compute the mean $\mathbb{E}_{\text{base}}[q]$ and standard deviation $\operatorname{std}_{\text{base}}[q]$ of the observable $q$ within this window to estimate the numerical noise floor.
        
            \item \textbf{Noise Thresholds:} We define a detection threshold $T_{q} = \mathbb{E}_{\text{base}}[q] + 3\operatorname{std}_{\text{base}}[q]$, corresponding to a $3$-std deviation from the baseline noise. A similar threshold $T_{s}$ is defined for the internal spread (standard deviation) of the population distribution.
        
            \item \textbf{Transition Detection:} Scanning through the parameter space, a point $m_i$ is flagged as a candidate transition if either of the following conditions is met for a persistence of 10 consecutive steps:
            \begin{itemize}
                \item \textit{Signal Shift:} The lower bound of the distribution departs from the baseline, $ \mathbb{E}[q](m_i) - \operatorname{std}[q](m_i) > T_q$, where $\mathbb{E}[q](m_i)$ and $\operatorname{std}[q](m)$ are respectively the average and standard deviation of $q$ among the population.
                \item \textit{Variance Explosion:} The standard deviation of $q$ among the population exceeds the baseline fluctuation threshold, $\operatorname{std}[q](m_i) > T_{s}$.
            \end{itemize}
        
            \item \textbf{Interpolation and Error Estimation:} Once the transition interval $[m_{i-1}, m_i]$ is identified, we approximate the behavior of $q$ linearly between these points. The critical value $m_c$ is defined as the exact point where this linear interpolation intersects the detection threshold $T$. To estimate the uncertainty $\delta m_c$, we combine the grid resolution $\Delta m = m_i - m_{i-1}$ with the error propagated from the baseline noise. This definition ensures that sharp transitions have small errors dominated by the grid size, while slow, continuous transitions have larger uncertainties, reflecting the sensitivity of the critical point to numerical noise.
        \end{enumerate}
        
        This methodology adapts naturally to the different physical regimes of the system. In the topological case ($\sigma=0$), the transition is sharp: the cavity variance $q$ exhibits a sudden jump from zero to a finite value. Our algorithm detects this discontinuity immediately, as shown in \cref{fig:critical_example_topo}.
        
        Conversely, for the disorder-driven SE to ME transition ($\sigma > 0$), the change is slow and continuous. As illustrated in \cref{fig:critical_example_SE_ME} (for $K=10, \gamma=-1$), the distribution of cavity variances spreads slowly from a delta function at zero to a broadened distribution. Here, the algorithm identifies the first value of $m$ where the population statistics deviate sufficiently from the baseline, with the propagated error capturing the smoothness of the crossover.
        
        Finally, the nature of the instability is classified by monitoring the mean abundance $\mu$. If $\mu$ remains finite across the transition, it is classified as a transition to the Multiple Equilibria (ME) phase (\cref{fig:critical_example_SE_ME}). If instead $\mu$ diverges, it is classified as a transition to the Unbounded Growth (UG) phase, as shown in \cref{fig:critical_example_SE_UG}. This fully automated procedure ensures that the phase diagrams are derived without manual bias.
        
    \subsection{\label{supp:interpolation}Phase boundaries interpolation}
\begin{figure*}[t!]
            \centering
            \begin{minipage}{0.32\linewidth}
                \includegraphics[width=\linewidth]{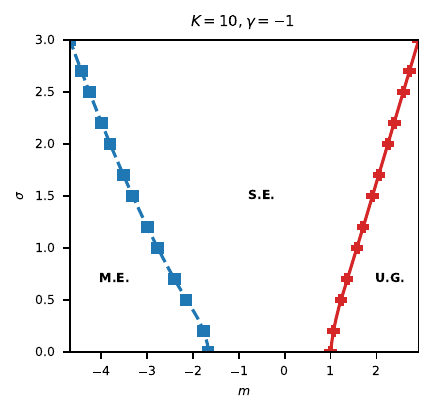}
            \end{minipage}
            \hfill
            \begin{minipage}{0.32\linewidth}
                \includegraphics[width=\linewidth]{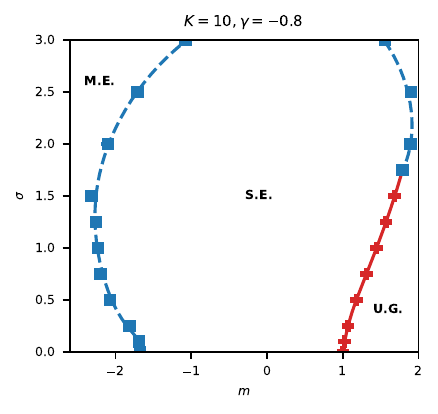}
            \end{minipage}
            \hfill
            \begin{minipage}{0.32\linewidth}
                \includegraphics[width=\linewidth]{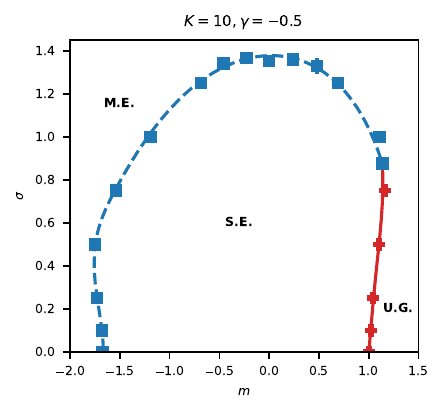}
            \end{minipage}
            \vspace{0.2em} 
            \begin{minipage}{0.32\linewidth}
                \includegraphics[width=\linewidth]{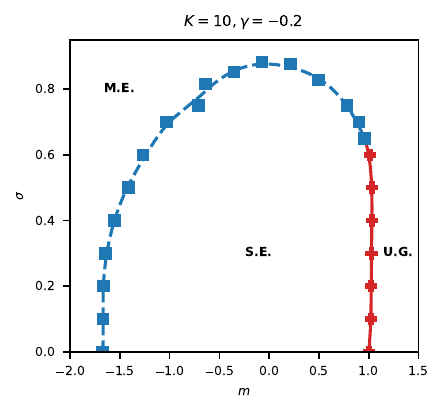}
            \end{minipage}
            \hfill
            \begin{minipage}{0.32\linewidth}
                \includegraphics[width=\linewidth]{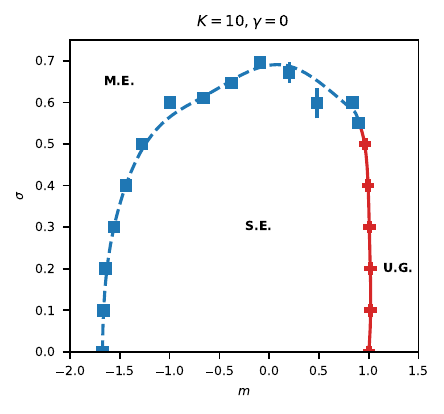}
            \end{minipage}
            \hfill
            \begin{minipage}{0.32\linewidth}
                \includegraphics[width=\linewidth]{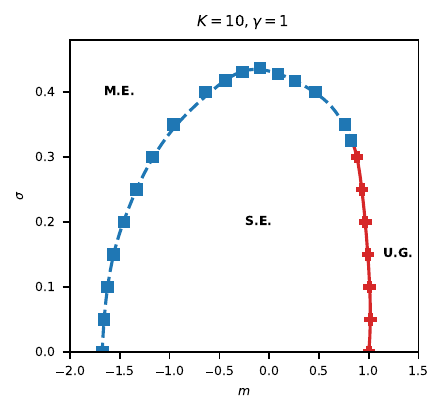}
            \end{minipage}
            \vspace{0.2em} 
            \begin{minipage}{0.32\linewidth}
                \includegraphics[width=\linewidth]{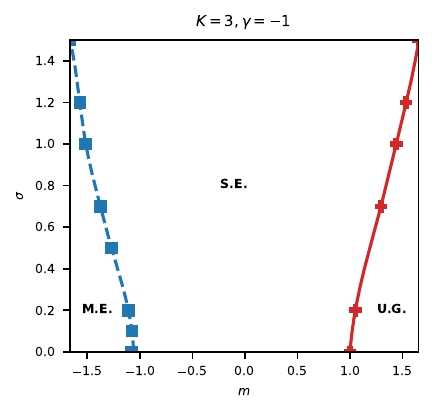}
            \end{minipage}
            \hfill
            \begin{minipage}{0.32\linewidth}
                \includegraphics[width=\linewidth]{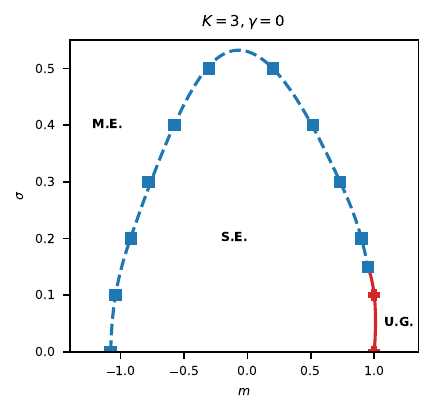}
            \end{minipage}
            \hfill
            \begin{minipage}{0.32\linewidth}
                \includegraphics[width=\linewidth]{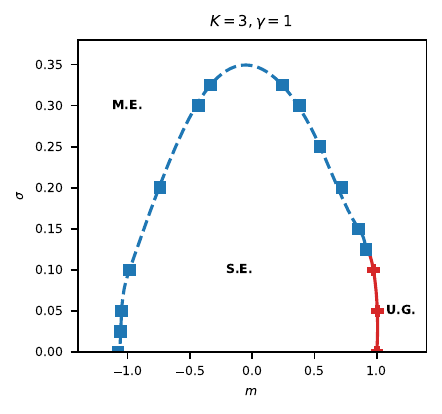}
            \end{minipage}
        
            \caption{\label{fig:phase_diagrams} \textbf{Interpolated stability phase diagrams.} The grid displays the individual stability boundaries computed for different network configurations. 
            \textbf{Top and Middle Rows ($K=10$):} Phase diagrams for $\gamma=-1.0, -0.8, -0.5$ (top) and $\gamma =-0.2, 0.0, 1.0$ (middle). 
            \textbf{Bottom Row ($K=3$):} Phase diagrams for $\gamma =-1.0, 0.0, 1.0$. 
            In each panel, the symbols represent the discrete critical points determined via the automated algorithm, while the lines represent the smooth parametric B-spline interpolation of order $k=3$. The regions correspond to Single Equilibrium (SE), Multiple Equilibria (ME), and Unbounded Growth (UG) phases.}
        \end{figure*}

        The stability phase diagrams presented in the main text (Fig. 3) and expanded here in \cref{fig:phase_diagrams} were constructed by mapping the critical boundaries in the $(m, \sigma)$ plane for fixed connectivity ($K=3, 10$) and interaction symmetry ($\gamma = -1, -0.8, -0.5, -0.2, 0, 1$).

        For each configuration $(K, \gamma)$, we performed a series of parameter sweeps to locate the discrete transition points $(m_c, \sigma_c)$ separating the Single Equilibrium (SE), Multiple Equilibria (ME), and Unbounded Growth (UG) phases. Depending on the local slope of the boundary, we either fixed $\sigma$ and scanned $m$ (for horizontal-like boundaries) or fixed $m$ and scanned $\sigma$ (for vertical-like boundaries).
        
        To obtain continuous phase boundaries from these discrete datasets, we performed a parametric interpolation using the Julia package \texttt{Dierckx.jl}. Specifically, we fitted a smooth parametric B-spline of order $k=3$ to the sequence of detected critical coordinates $\{(m_c^{(i)}, \sigma_c^{(i)})\}_i$. This procedure ensures a robust reconstruction of the stability limits, smoothing out small numerical fluctuations while preserving the global topology of the phase diagram.
        
%

\end{document}